\documentstyle[12pt,epsf]{article}

\newlength{\dinwidth}
\newlength{\dinmargin}
\def \MSbarbasic {\overline{{\rm MS}}}
\def \MSbar {\ifmmode \MSbarbasic \else $\MSbarbasic$\fi }
\def\lapproxeq{\lower .7ex\hbox{$\;\stackrel{\textstyle
<}{\sim}\;$}}
\def\gapproxeq{\lower .7ex\hbox{$\;\stackrel{\textstyle
>}{\sim}\;$}}
\def\be{\begin{equation}}
\def\ee{\end{equation}}
\def\bea{\begin{eqnarray}}
\def\eea{\end{eqnarray}}

\makeatletter
\def\fmslash{\@ifnextchar[{\fmsl@sh}{\fmsl@sh[0mu]}}
\def\fmsl@sh[#1]#2{%
\mathchoice
{\@fmsl@sh\displaystyle{#1}{#2}}%
{\@fmsl@sh\textstyle{#1}{#2}}%
{\@fmsl@sh\scriptstyle{#1}{#2}}%
{\@fmsl@sh\scriptscriptstyle{#1}{#2}}}
\def\@fmsl@sh#1#2#3{\m@th\ooalign{$\hfil#1\mkern#2/\hfil$\crcr$#1
#3$}}
\makeatother

\begin{document}
\titlepage
\begin{flushright}
DTP/97/78 \\
September 1997 \\
\end{flushright}

\renewcommand{\thefootnote}{\fnsymbol{footnote}}
\vspace*{2cm}
\begin{center}
{\Large \bf Charm in Deep Inelastic Scattering}\footnote[2]{Contribution to the Proceedings of the Madrid low $x$ Workshop, June 18-21 1997, Miraflores, Spain}
\renewcommand{\thefootnote}{\arabic{footnote}}

\vspace*{1cm}
J.C.\ Collins$^1$, A.D.\ Martin$^2$ and M.G.\ Ryskin$^{2,3}$

\vspace*{0.5cm}

\begin{tabbing}
$^1$xx\= \kill
\indent $^1$ \> Department of Physics, Pennsylvania State
University, University Park PA 16802, USA. \\
 
\indent $^2$ \> Department of Physics, University of Durham,
South Road, Durham, DH1 3LE, UK. \\

\indent $^3$ \> Petersburg Nuclear Physics Institute, 188350,
Gatchina, St. Petersburg, Russia.
\end{tabbing}
\end{center}
\vspace*{2cm}

\begin{abstract}
We outline the existing descriptions of the charm component of
the deep inelastic proton structure function $F_2$.  We discuss
recent approaches to include charm mass effects in the parton
evolution equations and the coefficient functions.
\end{abstract}

\newpage

Both H1 and ZEUS have used $c \rightarrow D^*$ to measure charm
production in the HERA domain \cite{DAUM}.  They find $F_2^c /
F_2 \sim 25 \%$, as compared to the EMC measurement of $F_2^c /
F_c \sim 1 \%$ at lower fixed-target energies.  Moreover, the
precision of the measurements of $F_2^c (x, Q^2)$ will improve,
particularly when vertex detectors become available.  Such
measurements should reveal important information on the gluon,
which enters at LO through the photon-gluon fusion (PGF) process
$\gamma g \rightarrow c \overline{c}$.  Clearly it is important
to see how best to include the charm mass effects in the
analysis. At NLO we have
\be 
\label{eq:a1}
F_2^c (x, Q^2) \; = \;  \int_x^1 \: dz \frac{x}{z}
\: \left[ \frac{8}{9} \; C_{q=c} (z, Q^2, \mu^2) c \left(\frac{x}{z}, \mu^2
\right) \: + \:  \frac{4}{9} \; C_g (z, Q^2,
\mu^2) g \left(\frac{x}{z}, \mu^2 \right) \right] ,
\label{F2}
\ee
where the coefficient functions $C_c = C_c^{(0)} + \alpha_S
C_c^{(1)}$ and $C_g = \alpha_S C_g^{(1)}$. \\

\noindent{\bf (a)~Massless charm evolution} \\

The simplest approach is to assume that the charm distribution
$c(x,\mu^2) = 0$ for $\mu^2 < \mu_c^2$, where $\mu_c \sim m_c$, and
then to evolve assuming $c$ is a massless parton both in the
splitting and coefficient functions.  This is the
approach used in the MRS and 
all but the latest (CTEQ4HQ \cite{LaiTung}) of 
the CTEQ global analyses.  MRS take
$\mu_c^2 = 2.7 {\rm GeV^2}$ so as to obtain a satisfactory
description of $F_2^c$ of EMC \cite{MRSA}.
On the other hand CTEQ set $\mu_c = m_c$; this is a consequence of their
choice to use the ACOT scheme \cite{ACOT2,ACOT1}, explained below, to
define the parton densities.  
In both cases, the number of quarks in the hard scattering coefficients 
is the number of ``active'' quarks.

Corresponding conditions are applied for the $b$ quark.  This is a
minor issue in the global fitting because of the small contribution of
processes involving $b$ quarks.

Although phenomenologically successful, the massless model
clearly is inadequate in the charm threshold region.  For
instance an on-shell $c \overline{c}$ pair can be created by
photon-gluon fusion (PGF) provided
$$
W^2 \; = \; Q^2 (1-x) / x \geq 4 m_c^2
$$
where $W$ is the $\gamma^* g \rightarrow c \overline{c}$ centre-of-mass
energy.  Thus at small $x$, $c \overline{c}$ production is not
forbidden even for $Q^2 < \mu_c^2$ where the massless approach
gives zero.  
The physical 
threshold for $c \overline{c}$ production, 
$W^2 = 4 m_c^2$, is not the threshold that is provided by the massless
model.  Only to the extent that charm production is small in the
threshold region does the massless model give a useful approximation
to $F_2$. \\

\noindent{\bf (b)~Photon-gluon fusion} \\

In this approach charm is treated as a heavy quark and not a
parton.  That is we put $c = 0$ for all $Q^2$ in (\ref{eq:a1})
and use the known $m_c \neq 0$ gluon coefficient function
$C_g^{\rm
PGF}$ for $C_g$.  This is called a fixed flavour number scheme
(FFNS) with $n_f = 3$.  In contrast to the situation for massless
quarks, there is no collinear divergence in $\gamma^* g
\rightarrow c \overline{c}$ since the integral over the $c
\overline{c}$ transverse momentum is regulated by $m_c$. 
However, this in turn means that at large $Q$ there are 
large logarithmic corrections in higher orders:
$F_2 \sim [\alpha_S \ln
(Q^2/m_c^2)]^n$ at $n^{\rm th}$ order.  In terms of the FFNS all
these contributions should be absorbed in the coefficient
function.  However, these are just the large logarithms that are
resummed\footnote{The NLO corrections to the PGF structure
function are known \cite{S1} and a leading twist analysis
\cite{S3} has been used to perform a resummation of the
$[\alpha_S
\ln (Q^2 / m_c^2)]^n$ terms for $Q^2 \gg m_c^2$.  Unfortunately
such an approach is not applicable for $F_2$ in the threshold
region $Q^2 \sim m_c^2$.} by DGLAP evolution in the purely massless
approach.  This implies that at large
$Q^2$ we should treat charm as a parton.  \\

\noindent{\bf (c)~Variable flavour number schemes:  ACOT and
MRRS} \\

The aim is to obtain a universal charm distribution $c (x, Q^2)$
all the way down to the resolution threshold $Q^2 \sim m_c^2$ in
order to make predictions for other processes.  To do this we
must include the nonzero mass of the charm quark in the calculation
in a consistent way.

The ACOT \cite{ACOT2,ACOT1} method is to exploit the techniques of
Collins, Wilczek and Zee \cite{CWZ}.  Below $\mu=m_c$, the parton
densities are those of the FFNS with $n_f=3$, and above $\mu=m_c$,
they are pure $\overline{{\rm MS}}$ distributions.\footnote{
    Note that the definition includes the full unapproximated
    dependence of the parton densities on the charm mass, and 
    that ``$\overline{{\rm MS}}$'' refers to the {\em ultra-violet}
    renormalization of the parton densities.  Also, the position of
    the change of definition need not be at $\mu=m_c$; any other value in
    the neighborhood would be suitable, provided that corresponding 
    changes in the matching conditions are made.
}
The evolution coefficients are independent of mass in the
$\overline{{\rm MS}}$ scheme, so they are the same as in a purely
massless calculation, with the number of active flavors changing from
3 to 4 at $\mu=m_c$.  
At leading order explicit calculation \cite{ACOT2} shows that the
parton densities are continuous at $\mu=m_c$.  The order $\alpha_S^2$
corrections to the matching conditions have been calculated \cite{S3}
but not yet implemented.  

In the ACOT scheme, the remaining dependence on the nonzero
charm-quark mass appears in the coefficient functions.  In the
threshold region, the charm density $c(x,\mu=Q)$ by itself does not
accurately provide the charm contribution to the structure function
$F_2^c$.  The necessary correction, at the first non-trivial order is
provided by the gluon coefficient in Eq.\ \ref{eq:a1}.  Consistency
with calculations in the FFNS is obtained at the appropriate level of
accuracy.

Buza et al.\ \cite{S3} have evaluated the coefficient functions
for $Q^2 \gg m_c^2$ in the VFNS and the FFNS to NLO.  
Their definition of the parton densities is exactly that of ACOT
\cite{S4}. 
They find
that for $Q^2 \gapproxeq 20 {\rm GeV}^2$ the VFN and FFN schemes agree
very closely.  They also work out the matrix
relation in the limit $Q^2 \gg m_c^2$ which allows a matching of
all the parton four flavour densities above the scale $ \mu^2 =
m_c^2$, to the three flavour densities below that point.  A direct
calculation of the matching conditions \cite{match-JCC} from the
partonic operator matrix elements confirms the matching conditions
computed by Buza et al., for the subset of the cases that have so far
been computed.  
This
relation may be used to calculate other inclusive cross sections
(e.g. large $E_T$ jets etc.)  This particular calculation only applies
for $Q^2 \gg m_c^2$, which does not solve the problem of what to do at
$Q^2 \sim m_c^2$.
One of the authors (JCC) considers the principles of the problem
solved; a paper is in preparation \cite{h-quark-fact}.
Work is in progress to obtain the ACOT coefficient functions at order
$\alpha_S^2$ from the Buza et al.\ calculations; these coefficient
functions will apply at all values of $Q$ of order $m_c$ and larger.
Another author (MGR) considers that
there has to be another matching condition which provides parton densities which are continuous in the threshold region.  

In common with the coefficient functions used in all the global
analyses, those of ACOT are for inclusive cross sections, $F_2$ for
the case under discussion here.  So a different, but related
calculation must be made if, for example, one wishes to compute the
distribution of charm in the final state.\footnote{
   A closely related case is that of the Drell-Yan cross-section,
   $d\sigma_{\rm DY} / dQ^2 dy$ integrated over transverse momentum
   $q_T$.  The physical cross section is analytic at $q_T=0$, but the
   lowest-order, or parton-model, contribution is proportional to
   $\delta(q_T)$.  Higher order terms give a cross section $d\sigma /
   d^4q$ that, without resummation, diverges as $q_T \to 0$ and that
   has delta-function with divergent coefficients.  Integration over
   $q_T$ cancels the divergences and gives a correct estimate of
   $d\sigma /dQ^2 dy$.}  The reason why it is necessary to treat each process separately can be illustrated by this example. The charm $p_T$ spectrum associated with the first $c \overline{c}$
pair produced by evolution from a gluon is proportional to
\be
\label{eq:z1}
\frac{d p_T^2}{p_T^2 + m_c^2} \; \left[ z^2 + (1 - z)^2 \; + \;
\frac{2 m_c^2 \; z (1 - z)}{p_T^2 + m_c^2} \right] .
\label{pt-spectrum}
\ee
This same $p_T$ spectrum appears in the calculation of the ACOT
matching conditions for the parton densities (integrated over $p_T$).
After application of the rules for computing an \MSbar{} distribution,
ACOT obtains the matching condition given previously.
However only the first term in eq.\ (\ref{pt-spectrum}) is operative in giving
the (mass-independent) evolution kernel of the ACOT distributions. The second term is hidden in the matching conditions and coefficient functions.
The full contribution to $F_2^c$ at order $\alpha_S$ also needs the
gluonic coefficient function in eq.\ (\ref{F2}).

The problem that worries MGR is that the second term in (\ref{eq:z1}) is proportional to $\alpha_S (m_c^2)$.  Thus it cannot be absorbed in a fixed order coefficient function,  which necessarily depends on $\alpha_S (Q^2)$.  On the other hand the matching takes place at one fixed point $Q^2 = m_c^2$, where we deal with $\alpha_S (m_c^2)$.  It is possible to choose such a matching which gives the correct description for $Q^2 \gg m_c^2$ but it looks impossible to correct the whole $p_T$ spectrum in the threshold region by just changing the parton densities at one point.

A different approach to include $m_c \neq 0$ effects has been
proposed by MRRS \cite{MRRS}.  Their aim is to formulate the
evolution procedure with $m_c \neq 0$ so as to generate universal
parton distributions applicable to any inclusive or exclusive
process with the (known) coefficient functions in the
conventional $\overline{{\rm MS}}$ renormalisation scheme.  In
order to do this
they analyse the leading log contributions which come from the
relevant (ladder type) Feynman diagrams.  It turns out that a
remarkable simplification occurs.  Recall that the leading logs
come from the configuration where the $p_T$ of the partons are
strongly ordered along the ladder.  At NLO accuracy it was found
to be sufficient to take into account the $m_c \neq 0$ effect
for the charm parton with $p_T \sim m_c$.  In fact all the $m_c
\neq 0$ effects at NLO occur only\footnote{Apart from the
corresponding
adjustment in the $\delta (1 - z)$ term in $P_{gg}$ \cite{MRRS}.}
in $P_{cg}$
\be
\label{eq:a3}
P_{cg} \; = \; \left[z^2 + (1 - z)^2 + \frac{2 m_c^2}{Q^2} \; z(1
- z) \right]
\ee
with $Q^2 = m_c^2 + p_T^2$.  In this way the MRRS procedure
automatically reproduces the correct $p_T$ spectrum of charm
quark.  It is straightforward to generalise the MRRS to higher
orders \cite{MRRS}. 

In a VFNS we have to take care to avoid double counting.  The
same Feynman graphs are generated by the evolution with
zero-order coefficient function (corresponding to $\gamma^* c
\rightarrow c$ with a spectator $\overline{c}$, or vice versa
with $c \leftrightarrow \overline{c}$) and the first order
coefficient function (which describes photon-gluon fusion $(g
\rightarrow c \overline{c}) \otimes (\gamma^* c \rightarrow c))$.
Thus we must subtract the contribution generated by evolution
from the PGF contribution.  This is done by both ACOT and MRRS.

The charm component $F_2^c$ of $F_2$ is totally determined by the
gluon and light quark densities.  The only parameter is the value
of $m_c$.  The MRRS predictions are shown in Fig.~1.  They are in
good
agreement with both HERA and fixed-target EMC data, which
together cover a
wide range of $x$ and $Q^2$.  Unfortunately in order to match on
to conventional $\overline{{\rm MS}}$ coefficient functions the
evolution scale is required to be $Q^2 = m_c^2 + p_T^2$ leading
to a rather low charm threshold at $Q^2 = m_c^2$. The value of
$\alpha_S$ is not sufficiently small in this region to neglect
NNLO corrections.  One of the main effects is to move
the threshold to $Q^2 \gapproxeq 4 m_c^2$ which, due to the
Uncertainty Principle, is the virtuality of the photon required
to resolve a charm quark within the $g \leftrightarrow c
\overline{c}$
fluctuations.  In the NLO predictions in Fig.~1 the NNLO
modification has been imposed by hand in a simplified way.  This
is the origin of the artificial structure at $Q^2 = 4 m_c^2$. 
Now that a complete NLLO calculation is available it may be used
to smooth out this behaviour. 

A corresponding plot from CTEQ, with the use of the ACOT scheme, is
shown in Fig.\ 2.  It compares the experimental data on the charm
contribution to DIS with calculations with two sets of parton
densities.  One set, CTEQ4M, is obtained from a global fit which uses
the massless scheme, and the other, CTEQ4HQ1, is obtained from a
global fit \cite{LaiTung} that used the ACOT scheme for DIS.

\newpage

\begin{figure}
   \begin{center}
      \leavevmode
      \epsfxsize=0.75\hsize
      \epsfbox{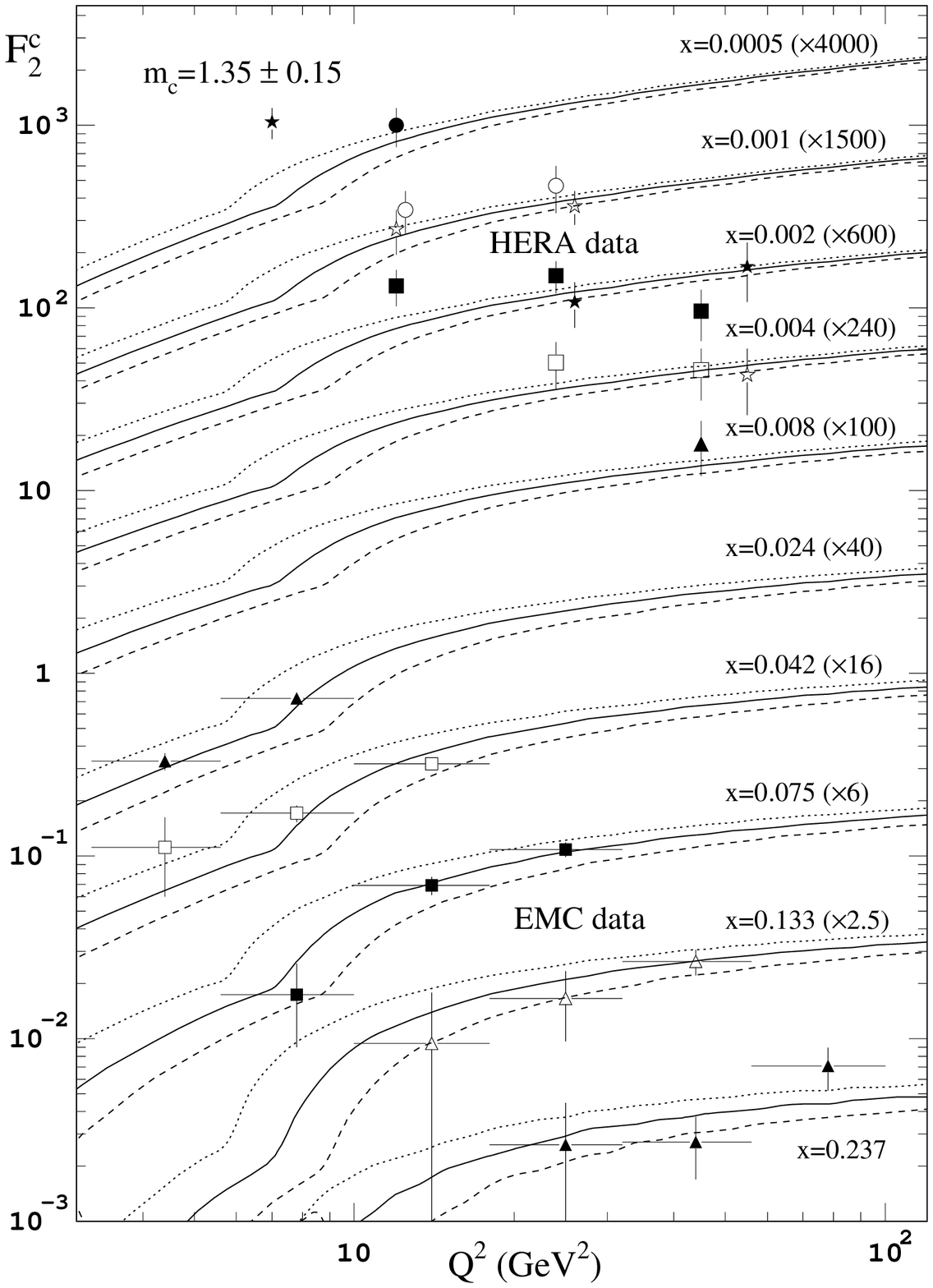}
   \end{center}
   \small{Fig.~1:  The predictions of MRRS \cite{MRRS} for $F_2^c$
compared with
the EMC \cite{EMC} and HERA \cite{DAUM} measurements.  The
dotted, continuous and dashed lines correspond to $m_c = 1.2, \;
1.35$ and 1.5 GeV respectively.  The starred data points in the
HERA domain are obtained by interpolating ZEUS measurements,
whereas the other HERA data correspond to the H1 measurements.}
\end{figure}

\begin{figure}
   \begin{center}
      \leavevmode
      \epsfxsize=0.8\hsize
      \epsfbox{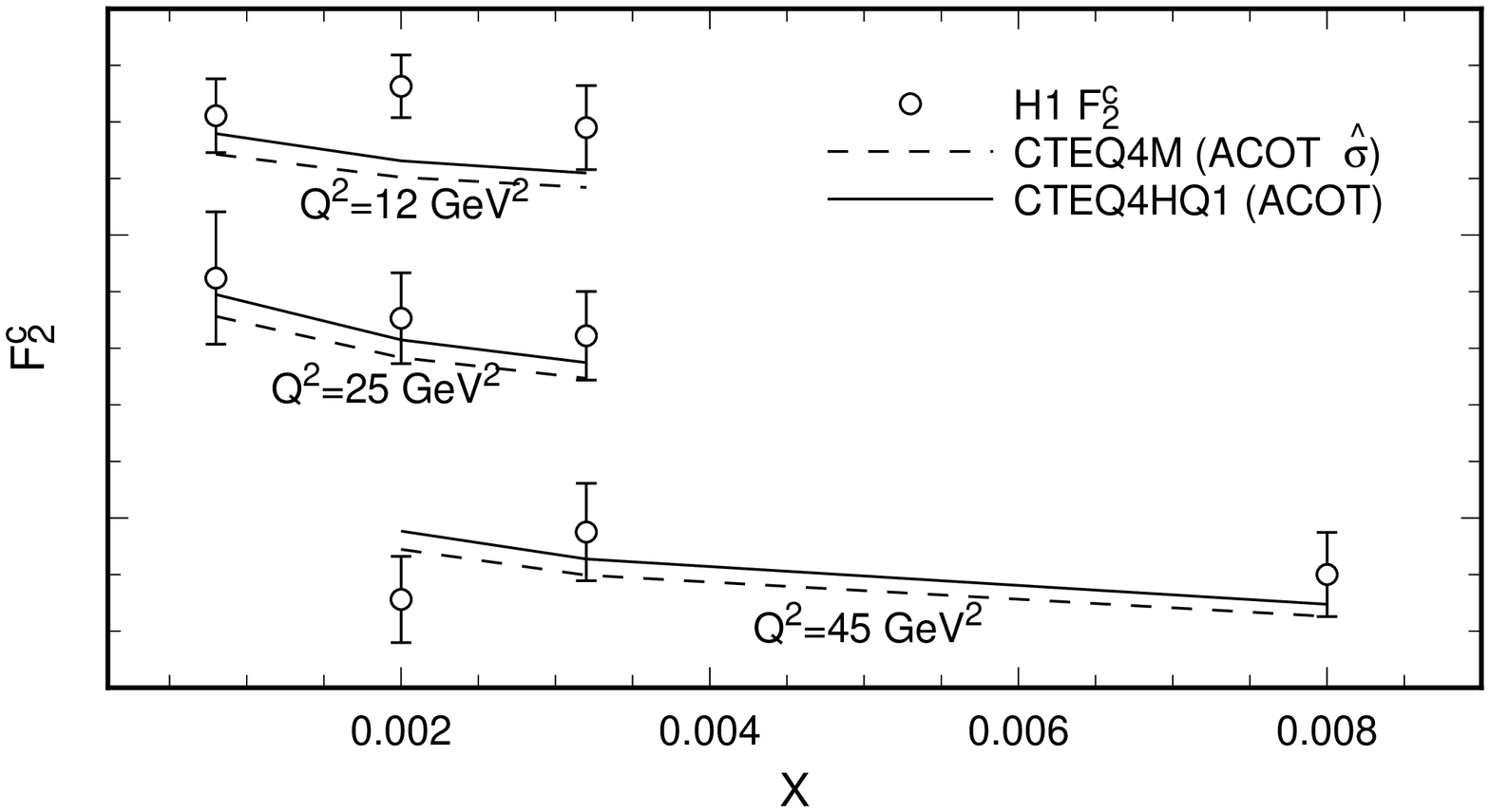}
   \end{center}
   \small{Fig.~2:  The CTEQ predictions \protect\cite{Tung}
   for $F_2^c$ when the ACOT scheme is used.  The dashed line
   gives the prediction from the CTEQ4M distributions, while the
   continuous line shows the prediction from the CTEQ4HQ1
   distributions, which were obtained from a global fit
   \protect\cite{LaiTung} where the ACOT method was used.  The data is
   from H1 \cite{DAUM}.
   }
\end{figure}

\end{document}